\documentclass[letterpaper, aps, tightenlines, 11pt, nofootinbib, preprintnumbers]{revtex4}
\pdfoutput = 1

\usepackage{setspace}
\usepackage[pdftex, hyperfootnotes = true, bookmarks = false, colorlinks = true, linkcolor = blue, citecolor = purple]{hyperref}
\usepackage{shortcuts}
\usepackage{titlesec}
\usepackage{xcolor}
\definecolor{Red}{rgb}{1,0,0}

\usepackage[margin = 2.5cm]{geometry}
\def\thesection{\arabic{section}}
\def\thesubsection{\arabic{subsection}}
\titleformat{\section}
  {\normalfont\Large\bfseries}{\thesection}{1em}{}
\titleformat{\subsection}
  {\normalfont\large\it}{\thesection.\thesubsection}{1em}{}
\titleformat{\subsubsection}
  {\normalfont\sl}{\thesection.\thesubsection.\thesubsubsection}{1em}{}

  \newcommand\ba{\begin{eqnarray}}
\newcommand\ea{\end{eqnarray}}
\newcommand{\bbb}{\ba\begin{array}{c}}
\newcommand{\eee}{\nonumber\end{array}\ea}
\newcommand{\een}[1]{\label{#1}\end{array}\ea}

\def\expp#1{\exp{\left ( {#1} \right ) }}
\usepackage{xcolor}
\definecolor{Red}{rgb}{1,0,0}
\definecolor{Blue}{rgb}{0,0,1}

  \newcommand{\IR}{\relax{\rm I\kern-.18em R}}

  \def\lll{_}
  \def\uu{^}
  \def\expp#1{\exp{\left ( {#1} \right )}}
  \def\sqd{^2}

\def\cc{\,}
\def\bha{\hat{\beta}}

\begin{document}

\begin{center}

\thispagestyle{empty}

\vspace*{5em}

{ \LARGE {\bf Topology of Future Infinity in dS/CFT}}

\vspace{1cm}

{\large Shamik Banerjee$^\bigstar$, Alexandre Belin$^\blacklozenge$, Simeon Hellerman$^\spadesuit$, Arnaud Lepage-Jutier$^\blacklozenge$,\\ Alexander Maloney$^\blacklozenge$, \DJ or\dj e Radi\v cevi\'c$^\bigstar$, Stephen Shenker$^\bigstar$}
\vspace{1em}

$^\bigstar$\ {\it Stanford Institute for Theoretical Physics and Department of Physics,\\ Stanford University, Stanford, CA 94305, USA} \\ \smallskip
$^\blacklozenge$\ {\it Physics Department, McGill University,\\ 3600 Rue University, Montr\'eal, QC H3A 2T8, Canada}\\ \smallskip
$^\spadesuit$\ {\it Kavli IPMU (WPI), The University of Tokyo Kashiwa, Chiba 277-8583, Japan}\\ 
\vspace{1em}
\texttt{bshamik@stanford.edu, alexandre.belin@mail.mcgill.ca, simeon.hellerman.1@gmail.com, arnaud.lepage-jutier@mail.mcgill.ca, maloney@physics.mcgill.ca, djordje@stanford.edu, sshenker@stanford.edu}\\
 \end{center}

\vspace{0.04\textheight}
\begin{abstract}

\begin{normalsize}
The dS/CFT proposal of Anninos, Hartman, and Strominger relates quantum Vasiliev gravity in dS$_4$ to a large $N$ vector theory in three dimensions.  We use this proposal to compute the Wheeler-de Witt wave function of a universe having a particular topology at future infinity. This amplitude is found to grow rapidly with the topological complexity of the spatial slice; this is due to the plethora of states of the Chern-Simons theory that is needed to impose the singlet constraint. Various mechanisms are considered which might ameliorate this growth, but none seems completely satisfactory.  We also study the topology dependence in Einstein gravity by computing the action of complex instantons; the wave function then depends on a choice of contour through the space of metrics.  The most natural contour prescription leads to a growth with genus similar to the one found in Vasiliev theory, albeit with a different power of Newton's constant.  
%prescription, based on analytic continuation from Euclidean AdS$_4$, appears to clash with certain reasonable assumptions about the physics of dS$_4$.  An alternate 

\end{normalsize}

\end{abstract}

\preprint{IPMU13-0123}

\maketitle

\pagebreak
\setcounter{page}{1}

\section{Introduction}

The quantum behavior of de Sitter (dS) space is deeply mysterious \cite{Anninos:2012qw}. This is in contrast with the behavior of anti de Sitter (AdS) space, which via the AdS/CFT correspondence is the best understood example of quantum gravity.  Not surprisingly, many authors have tried to use AdS/CFT ideas in dS space.  One  such proposal is the dS/CFT correspondence of Strominger \cite{Strominger:2001pn}.\footnote{
  A number of other proposals have been made with important new features. These include the causal patch representation \cite{Banks:2006rx}, the dS/dS  correspondence \cite{Alishahiha:2004md}, and the FRW/CFT \cite{Freivogel:2006xu} correspondence. We will not discuss these here.
} Closely related work was done by Witten \cite{Witten:2001kn}.   In Maldacena's interpretation of this proposal \cite{Maldacena:2002vr}, the partition function of the CFT --- regarded as a function of sources --- defines the Wheeler-de Witt (WdW) wave function as a function of bulk fields at future infinity. This will be the interpretation of interest in this paper.

Until recently, nothing was known about the CFT side of this duality, except for the properties determined from perturbative gravity calculations in the bulk.  An important step forward was taken by Anninos, Hartman, and Strominger, who formulated a precise proposal for the three-dimensional CFT dual to Vasiliev higher spin gravity in dS$_4$ \cite{Anninos:2011ui}. It had been previously proposed that even-spin and all-spin Vasiliev gravities in AdS$_4$  correspond to the CFTs of $N$ free scalar fields projected to the $O(N)$ or $U(N)$ singlet sector, respectively  \cite{Giombi:2012ms}; now, for the corresponding Vasiliev theories in dS$_4$, the CFTs are proposed to be theories of $N$ free Grassmann fields projected to the $Sp(N)$ or $U(N)$ singlet sectors, respectively. Similarly, critical (Wilson-Fisher) versions of these CFTs have been conjectured to be dual to Vasiliev gravities with a different choice of boundary conditions. 

In a further development, Anninos, Denef, and Harlow have computed the WdW wave function in the even-spin Vasiliev theory as a function of the uniform background bulk scalar field by computing the partition function of the Grassmann field theory as a function of the mass term \cite{Anninos:2012ft}. This allows one to make interesting statements/predictions about the Vasiliev dS theories using the above dS/CFT duality. Ref.~\cite{Anninos:2012ft} found that while zero mass is a local maximum of probability (i.e.~of the square of the wave function), there are other even higher maxima of probability at finite values of the scalar field background.  These other maxima seem to correspond to nonperturbative instabilities around the perturbative dS space solution.\footnote{This instability and possible restrictions on it have recently been discussed in \cite{Anninos:2013rza}.} Similar results were found in \cite{Castro:2012gc} for the wave function of pure three dimensional dS gravity, regarded as a function of the 
conformal 
structure at future infinity.  

In this paper we investigate the properties of Vasiliev gravity in dS$_4$ by looking at the dS/CFT proposal from a different angle. In Section \ref{VasilievSec}, we propose to study the WdW wave function when the future boundary has more complicated topology. The requirement that the boundary theory is in the singlet sector plays a crucial role here. This is because the only known way to project onto a singlet sector using a local, explicit Lagrangian formulation involves the weak coupling of vector matter to a Chern-Simons (CS) theory.  CS is a topological theory which develops additional states on spatial manifolds with a nontrivial fundamental group. We find evidence that these states lead to another instability --- one toward complicated boundary topology.  In particular, the WdW amplitude of finding a boundary with the topology of a genus-$g$ Riemann surface times a circle is found to be proportional to $k^{N^2 (g-1)}$.  Here $k$ is the (large) level of the associated CS
theory and $N$ scales as $1/G_N$, the inverse of the bulk Newton's constant. Thus, the wave function grows with the genus as $\exp\!\left[{1\over G^2_N}(g - 1) \ln k \right]$, revealing an $1/G_N^2$ instability effect in the bulk. In Section \ref{RemovalSec} we explore various mechanisms which might eliminate this exponential growth with genus, but we do not find a satisfying one.

It is natural to ask whether this exponential growth with genus is a generic feature of the quantum theory of de Sitter space, or a special feature of Vasiliev theory. To address this, we  begin in Section \ref{NonunitaritySec} with a general discussion of features of Euclidean conformal field theories holographically dual to dS  via a dS/CFT duality.  We obtain consistency conditions which encode the condition that the local physics in the bulk is that of gravity and matter satisfying the usual (bulk) unitarity conditions.  We argue that the partition function of a good dS/CFT dual will deviate from that of a unitary CFT only in a very particular way, while sharing certain exact properties with unitary partition functions in other respects. These requirements set even more stringent constraints on possible ways to remove large-genus divergences, and thereby they provide more justification for taking the large-genus effect seriously. 

We then discuss more general properties of the WdW wave function as a function of genus.  In Section \ref{EinsteinSec} we consider the case of  Einstein gravity with a positive cosmological constant.  In Einstein gravity there are field configurations that naturally interpolate  between different topologies (i.e.~cobordisms), and so it is natural to study the probability as a function of topology.\footnote{This was considered, for example, in \cite{Gibbons:1990ns}.}  We determine the WdW wave function in a semi-classical limit by computing the regularized action of complex solutions to the equations of motion (instantons).  The genus dependence  of the wave function depends in detail on precisely which solutions contribute.  In principle, one can determine which solutions contribute by defining an appropriate contour through field space.  In practice, one does not know how to define this contour and some sort of {\it ad hoc} prescription must be used.  We consider two natural prescriptions. The first is 
simply to 
include all saddles. This leads to a probability that increases exponentially with topology, as in Vasiliev gravity, but with  a different strength.   The probability of a genus-$g$ Riemann surface times a circle is found to be proportional to $\exp\!\left[{1\over G_N} (g-1)\right]$.  The other prescription we consider is to include only those saddles which arise in the analytic continuation from Euclidean AdS.\footnote{At the level of perturbation theory around global dS, this prescription works perfectly; it precisely reproduces the usual Hartle-Hawking (Bunch-Davies) wave function \cite{Maldacena:2002vr}.  Its validity at the non-perturbative level is unclear. }  This leads to a probability which vanishes exponentially with genus, in tension with the Vasiliev result.  Moreover, the contour leads to certain other features which seem to conflict with the ones considered in Section \ref{NonunitaritySec}.  We take this as evidence against the contour defined by analytic continuation from Euclidean AdS, and as 
an indication that the genus divergence occurs even in Einstein gravity.

We conclude with a discussion of open issues in Section \ref{DiscussionSec}.

\section{Vasiliev dS$_4$ and future topology} \label{VasilievSec}

In this section we study future topology in asymptotically dS$_4$ universes in Vasiliev gravity \cite{Giombi:2012ms}. Vasiliev gravity is a remarkable classical gauge theory that contains fields of arbitrarily high spin; in this sense it resembles a tension-free limit of a string theory. However, its Lagrangian formulation and quantization rules are currently unknown, and hence we cannot explicitly compute the WdW wave function in this theory. At present, the only reasonable way to understand quantum (or at least semiclassical) Vasiliev dS is through the holographic prescriptions of Strominger and Maldacena. This is the tool we will exclusively use below. 

\subsection{The divergence at large genus}

The issues we wish to highlight can be illustrated already in the context of the AdS/CFT correspondence, which posits a duality between 4D Vasiliev gravity in AdS and the singlet sector of the 3D $O(N)$ or $U(N)$ vector model at a critical point \cite{Giombi:2012ms}. Such a boundary theory is typically given a local, Lagrangian formulation by weakly coupling the critical  matter (taken to be scalar here) to a Chern-Simons (CS) topological field theory. The resulting action is
\bel{
  S =  \frac {ik}{4\pi}\int \Tr\left(A \wedge \d A + \frac23 A^3\right) + \int \d^3 x \sqrt{-g}\left(\frac12 D_\mu \phi^n  D^\mu \phi^n +  V\left(\phi^2\right)\right),
}
where $k$ is the CS level, $A$ is the gauge field one-form, $D_\mu$ is the gauge-covariant derivative, $\phi$ is a scalar field that transforms as an  $N$-component vector with $\phi^2 = \phi^n \phi^n$, and $V\left(\phi^2\right)$ is chosen so that the theory is conformal. (Different choices of this potential correspond to choosing the theory to be a marginal $\phi^6$ theory or to be at the Wilson-Fisher fixed point --- the critical theory.) On spacetimes of topology $S^1\times S^2$ the duality between this singlet theory and the Vasiliev bulk appears to work due to the following desirable features:
\begin{enumerate}
 \item The CS term introduces no additional gluon states, and yet (in the weak coupling/infinite level limit) it enforces Gauss' law that projects the matter spectrum onto just the singlets. Since the ``pure glue'' sector has only one state on this topology, the spectrum consists of only singlet states of matter fields, and these precisely match up with bulk Vasiliev fields, as described in the original proposal by Klebanov and Polyakov \cite{Klebanov:2002ja}. 
 \item CS-matter theories possess families of fixed points (indexed by the  't Hooft coupling $\lambda = N/k$ and distinguished by different matter potentials $V(\phi)$) that map onto the known families of Vasiliev theories \cite{Aharony:2012nh}.
 \item The boundary correlation functions precisely match the AdS/CFT correlators from Vasiliev gravity when the boundary is $\IR^3$ \cite{Giombi:2009wh}. This is a consequence of the presence of higher spin symmetries in the theory \cite{Maldacena:2011jn, Maldacena:2012sf}.
\end{enumerate}

Generalizations of the two dualities above to other topologies are still far from  fully understood. For instance, a puzzle is raised by the observation that a CS-matter system on $S^1\times T^2$ contains additional light states stemming from nontrivial holonomies along the cycles of $T^2$ \cite{Banerjee:2012gh}. These light states do not have known duals in Vasiliev theory.  They are closely related to the light states found in 2D CFT duals of higher spin gravity in $AdS_3$ \cite{Gaberdiel:2012uj}.  

The light states are small deformations of the zero energy states in the pure CS theory.  Roughly speaking, the coupling between the CS sector and the matter sector is $1/k$ and so at large $k$ the CS states continue to be light and their entropy remains unaffected.  This was analyzed in  \cite{Banerjee:2012gh}.  On  $S^1 \times T^2$ at large $k$  the states of the pure CS theory  can be understood from the semiclassical quantization of the classical phase space of the CS theory, which is the ($2N$-dimensional) space of flat connections on $T^2$ \cite{Elitzur:1989nr}.    In the Wilson-Fisher fixed point theory (i.e.~the critical theory) the scalar sector is gapped and can be integrated out to give an effective potential for the flat connections, giving small gaps for the CS states which vanish when $k \rightarrow \infty$.  The decoupling of the matter is manifest.  At temperatures $T$ of order one,  all the CS states will be counted in the entropy of the CS-matter theory.  Except when otherwise noted, we will henceforward take $T$ (determined by the size of the $S^1$) to be in this range.  (The case of the free scalar theory is more subtle because of the scalar zero mode.  Nonetheless, as argued in  \cite{Banerjee:2012gh}, a similar picture holds.)    The partition function is 
\bel{
  Z_{S^1 \times T^2} \sim k^N.
}
The scaling of this result follows from the dimension of the phase space and the role of $k$ as $1/\hbar$ in the semiclassical quantization.

These observations carry over to CS-matter theories on more complicated topology  \cite{Banerjee:2012gh}.  In this paper we will focus on $S^1 \times \Sigma_g$.    Here we need to take the critical theory to avoid the instability due to the $R \phi^2$ term in the action with $R$ negative.\footnote{This instability is cut off by the $\phi^6$ term, but at a value of $\phi$ that grows with $k$.} The value of the pure CS partition function is discussed in \cite{Witten:1988hf, Moore:1989yh, Marino:2011nm}.  The scaling can be determined by the dimension of the phase space.   There are $2 g$ holonomies, one fundamental group constraint on the product of the holonomies, and one overall $SU(N)$ rotation to eliminate.  This leaves a $(2g-2) N^2$ dimensional classical phase space.  (The matter phase space is only order $N$ dimensional and so is negligible.)   Quantizing semiclassically gives 
\bel{
\label{eq:result}
  Z_{S^1 \times \Sigma_g} \sim k^{(g - 1)N^2}.
}

The partition function for  higher genus surfaces diverges exponentially fast in the genus with a rate that increases with $k$ and $N$, both large in the `t Hooft limit appropriate for weakly coupled gravity.\footnote{To be precise, computing the one-loop determinant in CS theory reveals a subleading term that makes the partition function scale as $e^{(g-1) N^2 \ln(1/\lambda)}$ for 't Hooft coupling $\lambda = N/k \leq 1$, but for simplicity we will continue talking about $k^{(g-1)N^2}$ scaling.} Note that the $N^2$ in the above exponent is formally of order $1/G_N^2$,  not the more conventional $1/G_N$ \cite{Banerjee:2012gh}.   This suggests that the Vasiliev theory should really be interpreted as the open string sector of an open-closed string theory where the $N^2$ degrees of freedom are part of the closed string sector \cite{Aharony:2011jz, Chang:2012kt, AHJVFuture}.

On the other hand, if the CFT is put on $S^3$, the partition function \emph{decreases} as \cite{Klebanov:2011gs}
\bel{
  Z_{S^3} \sim k^{-N^2/2}.
}
This was the first indication of the $N^2$ dependence in the CS-matter theory. On a lens space $S^3/\Z_p$ the decay is slower \cite{Radicevic:2012in}\footnote{
  The decay disappears at $p \sim N$, the regime in which the $S^3$ is so thinly ``squashed'' by the $\Z_p$ orbifolding of the Hopf fiber $S^1$ that it becomes similar to $S^1 \times S^2$ from the point of view of the CS theory. These decreasing magnitudes cannot be seen through direct canonical quantization, and hence they can not be associated with any ``state counting'' or ``negative entropy'' reasoning. The upshot of these calculations is that CS and CS-matter theories on three-spheres and lens spaces come with an inherently negative and large contribution to the free energy stemming from the gauge sector. },
\bel{
  Z_{S^3/\Z_p} \sim k^{-N^2/2p}.
}

The rapid increases of the partition function for large genus  persist in the Vasiliev context, to which we now turn. As mentioned in the introduction, Vasiliev gravity in dS can be studied through the non-trivial modification of the $O(N)$ version of this correspondence: when $N$ is even, the even-spin Vasiliev gravity in dS space appears to be dual to the $O(-N)$ vector model at a fixed point \cite{Strominger:2001pn, Anninos:2011ui, Anninos:2012ft}.\footnote{
  ``$O(-N)$'' is a convenient shorthand for ``$Sp(N)$ with all Young tableaux transposed.'' This shorthand is used because some results obtained from the $O(-N)$ singlet vector model are obtained from those in the $O(N)$ model by letting $N \mapsto -N$. However, not all calculable quantities of the two models are related in this way. For instance, the $Sp(N)$ group is non-compact while $O(N)$ is compact, so their volumes cannot be related by just a change of sign of $N$. Thus it might not be advisable to think of $O(-N)$ as a kind of analytic continuation from $O(N)$.\\
  \indent The $O(-N)$ statement of dS/CFT has been checked at the level of matching bulk and boundary correlators. It is of note that other gauge groups on the boundary could possibly lead to the same correlator matches; for instance, something like $USp(N)$ might provide a compact alternative to $Sp(N)$ that leads to equally acceptable correlators.\\
  \indent There exists a corresponding, as yet unchecked statement about the duality between the $U(-N)$ model and the all-spin Vasiliev gravity in dS, where ``$U(-N)$'' stands for ``$U(N)$ with all Young tableaux transposed.'' The matter Lagrangian in this case would be $D_\mu \chi^n D^\mu \bar\chi^n + V\left(\chi^n \bar \chi^n\right)$.  There are no compactness issues in this case.
} This CS-matter theory is given by the action
\bel{
  S =  \frac{ik}{4\pi} \int \Tr\left(A\wedge \d A + \frac23 A^3\right) + \int \d^3 x \sqrt{-g}\left(\frac12 \Omega_{nm} D_\mu \chi^n D^\mu \chi^m +  V\left (\chi^2\right)\right),
}
where $\Omega_{nm}$ is the canonical symplectic form, $\chi$ is a vector of $N$ anticommuting scalars (Grassman fields), and $\chi^2 = \Omega_{nm}\chi^n\chi^m$. It is assumed that, for a suitable choice of $V(\chi^2)$, this theory can be conformal in the large $N$ limit --- just like its $O(N)$ counterpart.

Under the proposed dS/CFT duality, the time direction in Vasiliev dS is holographically reconstructed, and the above CFT lives on the future infinity of the bulk. Moreover, the partition function of the boundary CFT is interpreted as a WdW wave function obtained in the semiclassical approximation. Thus the $O(-N)$ model CFT emerges as a potent tool for calculating probability amplitudes on the space of late-time fields in quantum Vasiliev gravity: one merely has to compute the modulus squared of the CFT partition function to find a proxy for the probability distribution on the space of manifolds at the future infinity of Vasiliev dS.\footnote{This number is merely a ``proxy'' because of the current lack of understanding of the measure on the space of future infinities. We will have little to say on the issue of the measure in this paper, and will continue to refer to the WdW amplitude as the ``probability.''} The large $N$ asymptotics (presented in this paper) give corresponding WdW amplitudes in the 
semiclassical approximation of Vasiliev gravity. 

The same asymptotics for the partition function discussed above for the CS-commuting scalar theory are present in the $U(-N)$  anticommuting scalar model CFT: $ ~Z_{S^1 \times \Sigma_g} \sim k^{(g - 1)N^2}$.  The dual of this CFT is conjectured to be Vasiliev all-spin gravity in dS space.   These asymptotics  follow because the $U(-N)$ model has the same gauge group as the $U(N)$ one, and, as in the commuting case,  the matter does not disturb the count of pure CS theory states at large $k$ and  finite $T \sim 1$.   The anticommuting nature of the fields means that a negative eigenvalue of the quadratic form in the action is not dangerous, and so either the free or critical theories are acceptable.  The asymptotic behavior of the $O(-N)$ CFT has never been worked out in detail, but behavior similar to that above must appear due to the existence of $\sim k^{(g-1)N^2}$ states found through semiclassical  quantization on $S^1\times \Sigma_g$. 

The $O(-N)$ theory, if constructed as an $Sp(N)$ singlet vector model, features an additional divergence due to the non-compactness of its gauge group. This is a curious and new issue, particularly because this divergence is associated to the even-spin Vasiliev theory but \emph{not} to the all-spin Vasiliev theory, and it can presumably be cured by focusing on a related compact group, say $USp(N)$. To avoid conceptual difficulties related to the non-compactness of $Sp(N)$, one may wish to think in terms of the $U(-N)$ vector model for the rest of this section. 

The rapid growth associated to higher-topology future infinities is intriguing because, as mentioned above,  Newton's constant in the bulk scales as $G_N \sim 1/N$; a growth exponential in $N^2$ signifies a $1/G_N^2$ effect in the bulk. This does not correspond to any known object in a gravity theory. It is currently unknown how one should approach these divergences or decays of the probability amplitude, and the attempt to understand these effects from direct computations in bulk Vasiliev gravity is beyond our scope. 

In Section \ref{RemovalSec} we will discuss several different paths one could take in the hope of understanding these issues by studying the boundary theory directly. Before doing so, we will turn to a few details of the higher spin dS/CFT construction which might effect our discussion.

\subsection{Fixed points after continuation}

The authors of \cite{Anninos:2013rza}  have raised several interesting issues about the persistence of the boundary field theory fixed points after continuing to the dS regime.  In particular, the fixed points found at finite $\lambda$ in \cite{Aharony:2011jz, Aharony:2012nh} might in principle cease to exist when $N \mapsto -N$.  On general grounds we expect the effect of matter to be a $1/N$ effect, and so do not expect the change $ N \mapsto -N$ to affect the leading large $N$ ``classical" fixed point.   The $1/N$ corrections are more subtle because of the existence of the almost marginal $\phi^6$ operator.  Nonetheless, we do not see a reason for the zero of the beta function to disappear.

This issue is more easily analyzed in another realization of vectorlike dS/CFT.   If we continue the free fermion realization of vectorlike AdS/CFT to dS we find the dual is a theory of  commuting $O(N)$ spin-half fields.   Here there is no almost marginal operator and hence it is manifest that  the CFT fixed point persists at finite $k$.   The CS coupling is discretized and so it does not run. We hope to return to this issue in the near future \cite{Anninos2TBA}.

The authors of \cite{Anninos:2013rza} have also raised the possibility that  the appropriate continuation to the dS regime may require $k \mapsto i k$ to ensure the proper reality conditions for the parity violating terms in the bulk Vasiliev theory.   We have examined a truncation of the full theory to spin-0 and spin-1 fields and do not find a need for this complexification. Nevertheless, a more complete analysis is called for.   

If (for some reason) imaginary $k$ is required, a whole host of problems would develop. The CS level would no longer be quantized and so it may run under RG. Monopole operators have $\Delta \sim k \sim N/\lambda$ and so under naive continuation have large imaginary $\Delta$.   This minimizes the classical instability due to highly irrelevant operators that will be pointed out in Section 4, but may  make the $1/N$ expansion, the standard $G_N$ expansion of quantum gravity, badly behaved.  In particular, consider the theory on $S^1 \times S^2$.  There are corrections to perturbation theory due to monopole ``states'' going  around the $S^1$ of order $\exp(-\Delta)$ where the radius of the $S^1$ is order one.  In AdS/CFT these are of order $\exp(-N/\lambda)$ and hence nonperturbatively small.  But if $k$ (and hence $\lambda$) are imaginary, then na\"ively these effects are of order $\exp(-i N/|\lambda|)$ and are of order one magnitude.   The $1/N$ series is no longer asymptotic.  Presumably a Stokes line has 
been 
crossed.\footnote{We thank Daniel Harlow for this observation.} 

\subsection{High temperature phases}

So far we have worked on manifolds $S^1 \times \Sigma_g$, with the size of $S^1$ taken to be of order one. CS-matter systems of the type considered in this paper possess high-temperature phases in which the contribution of matter to the free energy (of order $NT^2 \trm{Vol}(\Sigma_g)$) becomes comparable to the contribution of the CS sector, which is of order $N^2(g - 1)$ \cite{Shenker:2011zf}. By choosing to normalize the volume of the spatial manifold to $\trm{Vol}(\Sigma_g) = 2(g - 1)$, we find that the phase transition happens at temperatures $T \sim \sqrt N$. The partition function scalings that we have presented in the preceding are no longer valid in this case. It may be possible that the genus-dependent divergences are no longer present.

We will qualitatively analyze this situation and find that the genus enhancement is present at high temperatures --- moreover, no phase transition is found. After integrating out matter, the partition function of the $U(-N)$ model can be written as a matrix model. Let $\alpha_i$ be the eigenvalues of this matrix. The partition function will be of the schematic form (omitting constant prefactors and intricacies from multiple holonomies):
\bel{
  Z_{S^1 \times \Sigma_g} \sim \int  \exp\left\{-(g - 1)\sum_{i < j} \ln|\sin(\alpha_i - \alpha_j)| - T^2 (g - 1) \sum_i \cos\alpha_i \right\} \prod_i \d\alpha_i.
}
The first term in the potential for the $\alpha$'s comes from the gauge sector, while the second term comes from the matter sector.\footnote{The matter contribution receives an additional minus sign relative to Ref.~\cite{Shenker:2011zf} because the matter fields are anticommuting. This minus sign is, in fact, irrelevant; it merely determines the position around which eigenvalues will clump at the saddle point, but does not affect the existence of the phase transition. This is in contrast with models with adjoint matter.} The above integral can be analyzed by reducing it to a saddle point calculation at large $N$. The gauge sector potential aims to clump the eigenvalues together, while the matter sector aims to have each eigenvalue attain $\alpha_i = \pi$. At $T \sim \sqrt N$, these two criteria must be satisfied simultaneously, but this is possible, unlike in the case of $U(N)$ or $U(-N)$ models on $S^1\times S^2$ \cite{Shenker:2011zf, Anninos:2012ft}. Therefore, there is no phase transition in which the 
matter takes over the gauge sector's domination and qualitatively alters the behavior of eigenvalues. Rather, at the saddle point all eigenvalues clump at $\alpha_i = \pi$. In this regime, the partition function scales as
\bel{
  Z_{S^1 \times \Sigma_g} \sim e^{N T^2 (g - 1)},
}
and the large-genus divergence is still present. 

We have here given a very schematic derivation of the high-$T$ behavior of the partition function on $S^1 \times \Sigma_g$; it would be interesting to fill in the holes and develop a new matrix model that would correspond to the $U(-N)$ vector model at $g > 1$. This is yet another topic that is relegated to future work.

 %\subsection{Can the large-genus divergence be removed?}
\section{Can the large-genus divergence be removed?}
\label{RemovalSec}

Let us recap the analysis so far. We have seen that, in the higher spin dS/CFT proposal of \cite{Anninos:2011ui}, the probability that future infinity has topology $S^1 \times \Sigma_g$ diverges with genus as $k^{(g-1) N^2}$, where $k\to\infty$. The conservative and least speculative way to address this issue is to accept the dS/CFT dictionary but restrict the topology at future infinity.\footnote{We are indebted to Dio Anninos for numerous discussions on this subject.}  For example,  as stated above, dS/CFT does not run into trouble on topologies of the form $S^1\times S^2$. It is therefore possible to ask meaningful questions about which boundary geometry is picked out by the WdW wave function from the set of all topologically $S^1\times S^2$ manifolds. We may think of this restriction as a future boundary condition on the space of all possible dS universes. This avenue of research has led to interesting open questions concerning instabilities and remains relevant regardless of any topological 
considerations \cite{Anninos:2012ft, Anninos:2013rza}.  
Other topologies are amenable to the same treatment: one may compare the WdW amplitudes for different geometries that have the same topology. For instance, one might ask whether a Vasiliev spacetime prefers to  break spherical symmetry at future infinity while still retaining the topology of an $S^3$. Even if the dS/CFT prescription leads to diverging prefactors in path integrals, these would all cancel in the calculation of relevant relative probabilities.

However, it is easy to construct solutions to the classical equations of motion which asymptote to topologies of the form $S^1 \times \Sigma_g$ at future infinity.  These are quotients of dS$_4$ by a discrete subgroup of its isometry group.  As this solution is locally de Sitter, it clearly solves the Vasiliev equations of motion.  These solutions have singularities of Milne type in the far past, where presumably quantum corrections are needed to fully understand the geometry.  The existence of these solutions is certainly a necessary, though not sufficient, condition that the WdW wave function is supported on non-trivial topology.  It is likely that other, more complicated, solutions to the equations of motion exist with the same asymptotics. In the absence of an argument to the contrary, it seems unphysical --- and acausal --- to impose a restriction on the possible state of the universe in the far future.  We will therefore take all  dS/CFT results at face value. This means that high-genus surfaces at 
future infinity are preferred, and the wave function of Vasiliev de Sitter space is asymptotic form is an infinite-genus, ``future foam''-like structure.\footnote{We note that, even though the genus is divergent, the volume of future infinity is divergent as well.  Thus it may be  that an appropriately defined  topological density --- such as the number of handles per Hubble volume --- is finite.  
%Questions of whether this occurs can only be answered once an appropriate cutoff surface is introduced in the bulk We will not pursue this possibility here; to determine whether these infinities can cancel in this manner, one needs to construct the appropriate measure factors associated with eternal inflation.
}
%This kind of behavior is well-known in the studies of Einstein de Sitter quantum gravity [***REF***], and the fact that the Vasiliev de Sitter behaves in the same way (assuming the dS/CFT rules above) may be of significance in future studies that aim to connect Vasiliev and Einstein gravity. 
%On the other hand, it might happen that this coincidence is a red herring; the Einstein calculation reveals an $1/G_N$ effect while the Vasiliev calculation reveals an $1/G_N^2$ effect, so these might be completely different issues.

In the remainder of this section we will discuss whether this large genus divergence can be removed by some modification of the higher spin dS/CFT conjecture. In quantum field theory, we are used to the idea of using ``renormalization'' to remove infinite factors in the partition function.  It is natural to speculate that
the large $k$, large $N$, and large-genus divergences could be removed by something like a standard renormalization procedure.  However, not all divergences are created equal, and in this section we argue that the divergences of dS/CFT cannot be renormalized by any reasonable renormalization procedure.

\subsection{Local counterterms}

According to the usual quantum field-theoretic idea of renormalization, we are allowed to modify the action of a field theory by arbitrary local counterterms.  One might attempt to remove the $k^{(g-1)N^2}$ divergence of the partition function by adding a $k$-dependent local term to the action.  However, it is simple to check that there is no local term consistent with scale invariance that would accomplish this.  The pure CS action has no local operators of its own, which leaves only $c$-number terms built from the background metric.  
However, the only scale-invariant term which can be built from the background metric and its curvatures in three dimensions is the gravitational Chern-Simons term, which vanishes for geometries of $S^1 \times \Sigma_g$ topology. 
Terms involving the matter fields are already fixed by conformal invariance; there are no available counterterms in the matter sector. Therefore, there is no way to cancel the large $k$ divergence with operators constructed from the gauge connection and background metric. 

\subsection{Cancellation of divergences with local spectator fields}

We now explore more extreme modifications of the theory in order to remove the $k^{(g-1)N^2}$ divergence.  We first might imagine adding additional spectator degrees of freedom that are decoupled from the CS-matter system. We take these degrees of freedom to be local to retain some degree of predictivity and control.  This decoupled system could conceivably be chosen to have the effect of cancelling the large $k$ behavior.    The new degrees of freedom must form a  topological field theory in order to preserve the spectrum of the boundary theory on simple topologies. Hence, the only viable possibility  is another CS theory.\footnote{There exists one more topological field theory that can be put on a 3D manifold --- the BF theory. However, introducing the BF theory brings about other problems, as we will discuss below.}

Let us study the $U(-N)$ version of dS/CFT. It would be simplest if the large genus behavior could be removed by inserting a decoupled $U(N)$ CS sector into the boundary theory. However, this would clearly just exacerbate the divergences. To cancel the divergent prefactors of the form $k^{(g - 1)N^2}$, we would need to replace $k$ by $1/k$ or to replace $N$ by $iN$ in the counter-theory. Neither option, on its own, leads to a well-defined CS gauge theory: the CS level must be an integer to preserve invariance under large gauge transformations, and $N$ must be a real integer in order for the $U(N)$ algebra to close. Nevertheless, one may \emph{define} the desired theories through analytic continuation of well-defined CS theories \cite{Marino:2009dp, Witten:2010cx, Mkrtchyan:2013htk}. We conclude that one way to define a divergence-free boundary theory is to start with two CS gauge fields with the same level and gauge group, with only one of them coupled to matter, and then analytically continue the decoupled 
gauge  sector 
to either $1/k$ or $iN$. While this technically works, it cannot be implemented by a local Lagrangian because the state degeneracies, the coefficients of exponentials in along a thermal or timelike circle, cannot be integers.  

Even the weaker condition, that they have a Hamiltonian implementing translations along the circle direction,
cannot hold.  In the absence of a local definition this procedure is {\it ad hoc} and unpredictive.

\subsection{Explicit overall factor multiplying $Z\lll{\rm CFT}$}

One can take a different route and simply multiply the partition function by an overall coefficient.  For
a positive integer coefficient, there is a manifestly local way to do this: simply add a massive scalar field with multiple vacua at zero energy, and take the mass to infinity.  However, if we are interested in decreasing the value of the partition function rather than
increasing it, then we cannot do so with any sort of local or causal dynamics.

A prescription to multiply the partition function by an overall numerical coefficient is actually a special case of a set of spectator fields with arbitrarily nonlocal interactions. To see this, consider a decoupled real scalar spectator field $\alpha(x)$, and write the action
\begin{equation}
S[\alpha] \equiv \pi\, \int d^3 x\, \alpha(x)\, F(- \nabla^2)\, \alpha(x)\ ,
\label{ArbitraryNonlocal}
\end{equation}
where $F(\lambda)$ is an arbitrary function of its argument.  A path integral quantization of $\alpha$ with the usual measure gives the result
\begin{equation}
Z_{(\alpha{\rm ~fields})} = \prod\lll\lambda\, [F(\lambda)]^{\mp{1\over 2}}\ ,
\end{equation}
where $\lambda$ are the eigenvalues of the Laplacian and the sign $\pm$ depends on the statistics of $\alpha$.  By choosing
\begin{equation}
F(\lambda) = Z_0^{\mp 2}\,\delta_{\lambda, 0}\ ,
\label{FixedFreeEnergy}
\end{equation}
we can chooose to engineer the partition function to be equal to any positive value $Z_0$, regardless of the geometry or topology of the manifold, so long as it
is connected.  More generally, the partition function for $\alpha$ with this nonlocal action, will be $Z_0^{\# ({\rm components})}$ for a manifold with
several connected components.  Again, this procedure seems {\it ad hoc} and unpredictive.  In
particular, the action \eqref{ArbitraryNonlocal}, \eqref{FixedFreeEnergy} is completely
nonanalytic in derivatives and cannot be given a Hamiltonian representation or causal interpretation
with respect to any time direction.

\subsection{Cancellation of divergences with spectator boundaries}

Another {\it ad hoc} and unfounded way to renormalize to eliminate the $k^{(g-1)N^2}$ prefactors would require us to restrict our attention to certain bulk geometries with disconnected boundaries. To see why this might be of interest, notice that the $S^3$ CFT partition function decays as $k^{-N^2/2}$ while the higher genus partition function diverges as $k^{(g - 1)N^2}$. This means that a CS-matter theory on $(S^1\times \Sigma_g) \cup S^3 \cup \ldots \cup S^3$ (with $2g - 2$ three-spheres and with no matter living on any of them) would have no divergent prefactors in its path integral. The renormalization in this case consists of adding additional spheres to the boundary manifold instead of adding extra terms to the original boundary field theory. This constraint  is not satisfactory for a number of reasons.  We currently do not understand multi-component boundaries, and in particular we know of no superselection rules that would force a genus $g$ surface to be associated to any particular number of 
additional 
$S^3$ manifolds when they are boundaries of the same asymptotically dS universe.  In particular, there is no action of the type \eqref{ArbitraryNonlocal}, that can implement such a superselection rule.

\subsection{Yang-Mills and BF theories as projectors onto singlets}
 
We may also try to remove the $k^{(g-1)N^2}$ divergences by coupling the matter to a different gauge theory that would project into the singlet sector. We here explore both known alternatives. The obvious choice, a 3D Yang-Mills (YM) term $g\_{YM}^{-2}\,\Tr\left( F\wedge * F\right)$, would spoil the conformal invariance and add additional states in the adjoint representation, even when we take the coupling $g\_{YM}$ to be small \cite{Aharony:2003sx}. Moreover, calculating in the YM regime and letting $g\_{YM} \rar 0$ is a non-analytic limit. A simple way to see that something goes wrong is to rewrite the YM action by using a Lagrange multiplier:
\bel{
  \int[\d A]\ e^{-2g\_{YM}^{-2}\int\Tr(F\wedge * F)} =  \int [\d A\d B]\ e^{i\int \Tr(B\wedge F)\, -\, (g\_{YM}^2/2)\int \Tr(B\wedge * B)}.
}
Note that $B$ is locally a Lie algebra-valued one-form that transforms in the adjoint representation under the usual gauge transformations. The zero coupling limit leaves only $\L\_{BF} = \Tr(B\wedge F)$ in the Lagrangian. The gauge symmetry gets enhanced at $g = 0$, where we may shift $B \mapsto B + \d_A \Theta$ for any zero-form $\Theta$ (with $\d_A \Theta \equiv \d \Theta - i A\wedge \Theta$); the action will remain invariant due to the Bianchi identity. This huge local symmetry will make the partition function blow up unless we gauge it away; but gauging this symmetry is a non-analytic procedure, and we should not expect that the resulting gauge-fixed action will have anything to do with YM theory at small but non-zero coupling. This non-analyticity has been explicitly explored long ago in the context of 2D YM theories \cite{Witten:1991we, Blau:1993hj, Blau:1993tv}, but the result extends to three and more dimensions.

The only other option is to start directly from the zero-coupling version $\L\_{BF}$ of YM theory. This is the so-called BF theory \cite{Witten:1989sx, Horowitz:1989ng, Blau:1993hj, Baez:1999sr}. Na\"ively, this is a good idea: the $B$ field acts as a Lagrange multiplier that sets $F = 0$, any coupling can be eliminated by rescaling $B$ so there is no ``infinite level'' limit that might give rise to divergences, and the BF theory is topological and hence contains no extra local degrees of freedom. Moreover, unlike CS, BF theory can be formulated in any dimension, and so it seems like a natural candidate for a singlet-sector projector in any spacetime.

Canonical quantization shows that the BF theory does \emph{not} project into singlets, however. The Hilbert space of the theory is obtained from the space of solutions to the equations of motion  modulo gauge transformations. The equations of motion for the BF theory, coupled to a matter current $J$ through a $\Tr(A \wedge J)$ term, are
\bel{
  F = \d_A A = 0,\quad \d_A B = J.
}
In CS theory, taking the level to infinity enforced $J = 0$ by making all charged states infinitely heavy. On the other hand, the BF theory gives us no reason to set $J = 0$, and we do not achieve the desired projection. If we quantize these equations of motion, we will find charged states in the spectrum.

This is not all; the canonical quantization supplies us with more worries. In the CS theory, the path integral was normalized such that the partition function on $S^1\times S^2$ was equal to unity, i.e.~to the number of states in canonically quantized CS theory. In the BF theory, canonical quantization reveals an unpleasant but old feature \cite{Witten:1989sx, Baez:1999sr}: the phase space is non-compact (the gauge-invariant solutions to the equation of motion, $\d_A B = 0$, are classified by the non-compact cohomology group of $\d_A$). This means that the partition function on $S^1\times S^2$ must be infinite. At present, it is not understood how (or if) this infinity is to be regulated; consequently, the canonical normalization of the BF path integral is unknown.\footnote{The Abelian BF theory can be given a canonical interpretation when seen as a compact, $U(1)\times U(1)$ theory \cite{Witten:2003ya, Gukov:2004id, Hansson:2004}. This is no longer possible in a non-Abelian situation, unless we 
compactify certain modes of the gauge fields by introducing large
gauge transformations by hand.} Thus, BF theory is presently not a useful tool for analyzing singlet vector models.

\bigskip

Before leaving the subject of BF theory, let us address a tangential issue which may be of interest in any subsequent treatment of gauge theory path integrals of the kind appearing in this paper. This discussion also points out some common pitfalls one may face if trying to understand the above canonical treatment of BF theory using the path integral language. It is commonly said that the $B$ path integral sets $F = 0$. However, in general this is not precisely correct. We must fix both gauge invariances of the BF theory before we can understand the integral over $B$. Explicitly, the infinitesimal gauge transformations that are being fixed are
\bel{
  A \mapsto A + \d_A \Lambda,\quad B \mapsto B - i B\wedge \Lambda
}
and
\bel{
  A \mapsto A,\quad B \mapsto B + \d_A \Theta.
}
The presence of the second group of gauge transformations means that some of the $B$ field configurations must be equivalent, and hence we cannot path-integrate over $B$ without exercising utmost care.

To see where we stand, let us focus on spacetimes of topology $S^1 \times S^2$. Gauge fixing can force $A_0$ (the gauge field component along the $S^1$ direction) to be $x_0$-independent and to lie within the Cartan subalgebra \cite{Blau:1993hj, Blau:1993tv, 'tHooft:1981ht}. The elements $T^i$ of this subalgebra generate the maximal torus of the gauge group, and hence this choice may be called the ``torus gauge.'' This (with an $x_0$-independent Coulomb gauge, irrelevant for this discussion) fixes the first group of infinitesimal gauge transformations.

We are still left with the second group of transformations, the ones parametrized by the zero-form $\Theta$. These change $B$ but not $A$. Let us focus on shifts of $B_0$, the component of $B_\mu$ that points along the $S^1$ direction. These are given by
\bel{
  i\delta B_0(\omega, \b x) = \omega \Theta(\omega, \b x) + [A_0(\b x), \Theta(\omega, \b x)].
}
Note that we assume that $A$ is already gauge-fixed into the torus gauge. One advantage of this gauge is that the above transformations simplify greatly; the Cartan and non-Cartan components of $B_0$ transform as
\bel{
  i\delta B_0^i(\omega, \b x) = \omega \Theta^i(\omega, \b x),\quad i\delta B_0^\alpha(\omega, \b x) = \left(\omega +  \alpha_i A_0^i(\b x) \right) \Theta^\alpha(\omega, \b x).
}
The root vectors are defined by $[T^i, T^\alpha] = \alpha_i T^\alpha$ (no summation), and $\alpha$ indexes the non-Cartan generators $T^\alpha$. These transformations show that we may gauge-fix almost all modes of the $B_0$ field to zero. The only exceptions are $B_0^i(0, \b x)$ and those modes $B_0^\alpha(\omega, \b x)$ for which $\omega = - \alpha_i A_0^i(\b x)$. Moreover, we expect the latter set to be of measure zero when integrating over all $A_0^i(\b x)$, so we may really state that all modes of $B_0$ except for $B_0^i(0, \b x)$ can be gauge-fixed to zero.  In this temporal gauge the path integral over the $B$ field does \emph{not} set the $\omega\neq 0$ components of  $F^i_{12}(\omega, \b x)$ to zero, contrary to the usual statement that the integral over $B$ produces a $\delta$-function in $F$. All the other modes of $F$ are still set to zero, however.

We arrive at the following phenomenon: even though the canonical quantization shows that there are no $F \neq 0$ states, the torus gauge in the path integral forces us to perform a sum over some $F \neq 0$ sectors. These configurations are gauge artifacts, discovered in a very different form already in YM and CS theories \cite{Blau:1993hj, Blau:1993tv, 'tHooft:1981ht, Jain:2013py}. The point we would like to stress is that these seem to be a generic feature in gauge theories, and they need to be taken into account properly when path-integrating. The current lack of understanding of the BF path integral normalization forces us to relegate the study of these artifacts (and of the BF-matter path integral) to future work. 

\bigskip

The failures of the YM/BF theories as means to remove
the states of the Chern-Simons sector are, at any rate, a foregone conclusion in the unitary
case.  If this could be done, then for a $T\uu 3$ geometry the YM/BF + CS + matter system would amount to a projection of the matter theory onto its singlet sector without the addition of new states.  Such a projection is disallowed by a general no-go theorem in \cite{Banerjee:2012gh}, based on modular invariance.  In the nonunitary case relevant to dS/CFT, we do not yet have sufficient understanding of the basic rules of the CFT to formulate a no-go
result.  We will turn to the question of what rule might replace unitarity for dS/CFT, in Section \ref{NonunitaritySec}.

\section{Consistency conditions constraining the CFT of dS/CFT}
\label{NonunitaritySec}

We have so far considered various  {\it ad hoc} attempts to remove the large-genus divergence.  At this point the most we can say about them is that they seem rather contrived.  We will now try to understand more systematically what consistency conditions are expected for a CFT holographically dual to gravity with a positive cosmological constant. This way we may arrive at a set of principles to rule out --- or rule in --- various possible prescriptions.

As a first example we note that  
 this CFT must have an unusual property. Going to late time in dS corresponds to flowing to the UV in the boundary RG.   This means that if the CFT contains any (single trace) irrelevant operators, it has a classical instability at late times.  Thus a stable theory must have the highly unusual property of having no single trace irrelevant operators.\footnote{The monopole operators in the CS-matter theories at large $k$  discussed in Section 2.2 have $\Delta \sim k$ and so are highly irrelevant, at least for  positive $k$, which seems to indicate a strong classical instability. 
} 
%  In the bulk such an operator corresponds to a bulk tachyonic field.

In the case of AdS/CFT, the key principle from which various universal principles
and no-go theorems can be derived, is unitarity (see for instance \cite{Hellerman:2009bu}).  It is
not clear what plays the rule of unitarity in the dS/CFT context.
It is well known that a theory dual to bulk gravity {\it via} the dS/CFT correspondence cannot satisfy
the condition of unitarity (or reflection positivity) in the usual sense.  In the limit where the 
gravitational theory is described by Einstein gravity coupled to local effective field theory,
the spectrum of single-trace operators in
the CFT should correspond to the spectrum of single particle states \cite{Strominger:2001pn}:
% STARTT FIRST CHANGES MARKED IN BLUE % HEER

\bel{
E^{\rm CFT} = {1\over{r_{{\rm boundary~}S^2}}} \left [
{3\over 2} \pm \sqrt{{9\over 4} - \ell_{dS}^2  m^2} \right ]
\label{LowDimSpectrum}
}
for scalar primaries, and more generally
\bel{
E^{\rm CFT} = {1\over{r_{{\rm boundary~}S^2}}} ~ \Delta(m\sqd\ell_{dS}\sqd,J)
\label{LowDimSpectrumSpinJ}
}
for primary states with spin $J$ and mass $m$, with formul\ae ~$\Delta(m\sqd\ell_{dS}\sqd,J)$ given by analytically continuing the corresponding formul\ae\ for AdS by $\ell\lll{dS}\sqd m\sqd \to - \ell\sqd_{AdS} m\sqd$ (see e.g. \cite{Bousso:2001mw} for a discussion of this analytic continuation):
\bel{
\Delta(m\sqd\ell\sqd, J) \equiv \Delta^{[AdS]}(- m\sqd\ell\sqd_{AdS}, J)\ . 
}
The formul\ae~$\Delta^{[AdS]}(- m\sqd\ell\sqd_{\rm AdS}, J)\ $ are given for some low spin $J$, \it e.g., \rm in  \cite{Aharony:1999ti}.  In general, the dS formul\ae\ go as
  \bel{
  \Delta(m\sqd\ell_{dS}\sqd, J) \simeq {{\pm i \, m\ell_{dS} + O\left((m\ell_{dS})^0\right)}},
  }
at large $m$ and any fixed $J$. 

% ENDD FIRST CHANGES MARKED IN BLUE % HEER
From
this dictionary one can see that the spectrum of the dilatation operator is not generally real, except in special cases such as that of Vasiliev gravity \cite{Anninos:2011ui}. In this section we would like to distinguish several distinct ways in which a CFT (or a quantum theory more generally) can fail to be unitary, and to point out that these types of nonunitarity have differing logical status in the context of the
dS/CFT correspondence.

\subsection{Types of nonunitarity}

The Euclidean path integral of a unitary quantum theory on a circle of length $\beta$ represents
the thermal partition function
\bel{
Z(\beta) = \sum_n \expp{- \beta E_n}\ , \hskip1in
E_0 \leq E_n \in \IR\ .
\label{UnitaryForm}
} %end\bel
One can read off the spectrum
by going to a limit in which the length $\beta$ is much larger than the typical energy gap
in the spectrum $\{ E_n \}$, which in a CFT is the inverse length scale of the
spatial slice.

There are several ways in which the partition function of a quantum theory can differ from
the form \eqref{UnitaryForm}:
\begin{itemize}
\item{The system can display {\bf complex-enegy nonunitarity}:  The system may have exactly
the form \eqref{UnitaryForm}, but with $E_ n$ not all real.  The sum is still well-defined if
the real parts of $E_n$ are bounded below:
\bel{
Z(\beta) = \sum_n \expp{- \beta E_n}\ , \hskip1in
{\rm Re}(E_0) \leq {\rm Re}(E_n)\ .
\label{ComplexEnergyNU}
} %end\bel
}
\item{The system can display {\bf spectral-density nonunitarity}: The system may have the form
\bel{
Z(\beta) = \sum_n a_n~ \expp{- \beta E_n}\ ,
\label{SpecDensNU}
} %end\bel
with some or all of the $a_n$ not being positive integers.
}
\item{The system can display {\bf unbounded-energy nonunitarity}, where
the real parts of the $E_n$ are unbounded below.  The partition function does not converge in
this case.}
\item{The system can display {\bf continuum nonunitarity}, in which the set of 
$E_n$ is not discrete.  The partition function generally diverges in this case as well.}
\end{itemize}
A generic nonunitary theory can of course display several of these types of nonunitarity at once.

Continuum-type nonunitarity can be considered the mildest form of nonunitarity because it can occur
as a limiting case of unitary CFT, when a set of energy levels becomes increasingly
dense and collapses to a continuum in the limit, as exemplified 
by the $\epsilon\to 0$ limit of the
spectrum
\bel{
E_n \simeq E_0 + {{\epsilon}\over{n^2}},  
}
in some family of CFTs parametrized by $\epsilon$.  This type of nonunitarity occurs in very simple
families of CFTs, e.g.~in the moduli space of $c=1$ conformal field theories, with the radius $R$ of the target-space circle given by $R = \sqrt{\alpha^\prime/ \epsilon}$. This type of nonunitarity also arises in moduli spaces of strongly coupled unitary CFT as well, including theories holographically dual via AdS/CFT \cite{Seiberg:1999xz} to known superstring backgrounds.

None of the three remaining types of nonunitarity can arise as limits of unitary CFT with discrete spectrum.

\subsection{Unitarity properties of dS/CFT in the Einstein-gravity limit}

Having distinguished these possibilities, one is led to ask which of these types of nonunitarity can occur in holographic realizations of bulk quantum gravity. In the case of AdS/CFT, the 
global timeline Killing vector longtudinal to the boundary in Lorentzian signature allows us
to conclude straightforwardly that energies are real and the spectral density is positive, insofar
as the bulk description can be trusted.  In the dS/CFT context, one can also deduce certain unitarity properties of the CFT directly from the existence of a bulk dual.

By virtue of formul\ae\, \eqref{LowDimSpectrum}, \eqref{LowDimSpectrumSpinJ}, a CFT with an interpretation in terms of bulk particles in dS must necessarily display complex-energy nonunitarity. 
Here we make a complementary observation: The same requirement of a consistent bulk particle interpretation appears to forbid the unbounded-energy and spectral-density types of nonunitarity in the thermal
 partition function on $S\uu 2$ spatial slices of the boundary CFT.   
 
 First, the boundedness-below of the energy spectrum follows directly from the form of the expressions \eqref{LowDimSpectrum},  \eqref{LowDimSpectrumSpinJ},
which forces
the real parts of the energies to lie above zero. 
The boundedness-below of the energy is simplest to see for boundaries $\IR\times S\uu 2$, but the same principle applies to boundaries $\IR\times X_2$, where the spatial slices $X_2$ have topology other than $S\uu 2$, or possibly $S\uu 2$ topology with a non-round metric.  For boundary geometries $S\uu 1 \times X_2$, the eigenvalues of time translation along the $S\uu 1$ are
no longer determined directly by representation theory as they are for the round $S\uu 2$, but we can nonetheless rule out the possibility of unbounded-energy nonunitarity by considering the behavior of solutions to the wave equation.  For particle
states in the bulk with wavelengths $1/k$ much smaller than $\ell$, the global geometry and topology of $X_2$ are irrelevant, and the wavefunction of a particle sees only a local dS geometry at leading order in the short-wavelength/WKB/geometric optics approximation.    In this limit, the solution to the wave equation gives
\bel{
  E \simeq {1\over{r_{{\rm spatial~slice}}}}~\left [  i  \,k\ell+  c_0  + O\left(\frac1{k \ell}\right)
  \right ]\ ,
}
where $c_0$ depends only on the geometry of $X_2$
% STARTT SECOND CHANGES MARKED IN BLUE % HEER
{
 and the spin $J$ of the primary (replacing
${3\over 2}$ in the case of the round $S\uu 2$ and $J=0$).
}
% ENDD SECOND CHANGES MARKED IN BLUE % HEER
In particular, for any $\epsilon > 0$ there will be at most a finite number of states with ${\rm Re}(E)  < c_0 - \epsilon$.  The CFT does not display unbounded-energy nonunitarity for any geometry of its spatial slices, insofar as the Einstein-gravity approximation is reliable.

 The spectral-density type of nonunitarity also appears to be forbidden by a bulk particle interpretation of any kind, if the low-energy local physics in the bulk is described by a unitary Lagrangian such as Einstein gravity.  To see this, let us consider
the thermal partition function given by the path integral of de Sitter gravity on a bulk space-time with boundary $S\uu 1 \times S\uu 2$.  At low temperature, the dominant bulk solution is a quotient of global dS by a finite scale transformation.   In the regime where the dS radius is large compared to the Planck scale, the partition function is described to leading approximation by a path integral over the worldlines of free gravitons, scalar fields, and other particles propagating in this spacetime.\footnote{ The locally de Sitter geometry which asymptotes to $S^1 \times S^2$ --- or indeed $S^1 \times \Sigma$ for any Riemann surface $\Sigma$ --- has a spacelike singularity of Milne type.  This is a big bang singularity, at which one must fix some boundary conditions in order to compute the partition function.  We do not expect this choice of vacuum state to effect the 
positivity of the spectral density.}  
 The spectral density is automatically positive-integer and inherited from a description of the system in terms of real particles, as opposed to ghost-like particles.

 For spatial slices of any geometry, we conclude the partition function is of the form \eqref{ComplexEnergyNU}, insofar as the bulk description of the CFT is reliable.

\subsection{General unitarity rules for dS/CFT?}

We would of course like to learn about the unitarity properties of dS/CFT beyond the
limit where bulk physics of Einstein gravity coupled to local fields is the controlling approximation.
We observe here that dS/CFT displays only a very limited type of nonunitarity --- complex-energy
nonunitarity but not unbounded-energy or spectral-density nonunitarities --- not only in the Einstein-gravity limit but in the Vasiliev theory as well, which is in a drastically different range
of anomalous dimensions for the CFT.  The Vasiliev theory (assuming the correctness of the higher spin dS/CFT duality) appears to have the properties of positive spectral density and energy bounded below even at finite $N$, indicating that these are features of the full quantum theory.

Furthermore, in the event that one
considers families of dS/CFT theories that interpolate continuously between large and small coupling in the CFT (as defined, say, by the sizes of anomalous dimensions), 
an infinitesimal variation of the parameter must be realized as an operator insertion in the CFT.  Such an insertion can only alter the Hamiltonian of the CFT, but cannot continuously change the norm
of a state from $+1$ to $-1$.  Therefore a dS/CFT theory connected to the Einstein gravity limit
by membership in a continuous family must have positive integer spectral density, despite
the absence of a directly controlled particle interpretation in the bulk.

The form \eqref{ComplexEnergyNU} can thus be inferred for many CFTs realizing gravity through a holographic correspondence.  We would like to propose that it may be a universal rule for dS/CFT theories, a possible foundational principle to play part of the role of unitarity of the CFT in the AdS/CFT correspondence.  If true, the form \eqref{ComplexEnergyNU} would exclude the exotic genus-divergence-cancelling proposals discussed in Section \ref{RemovalSec}. 

\subsection{High-temperature behavior}

We would also like to comment on the relationship between  the form \eqref{ComplexEnergyNU},
and the high-temperature behavior of the CFT.  If we can assume the spectrum of the CFT does indeed have a partition function of the form \eqref{ComplexEnergyNU}, even in the high-temperature phase, then possible exotic 
high-temperature behaviors of the CFT partition function become easier to constrain. 

Reading off the single particle spectrum for the space-time with future boundary geometry $\IR \times S\uu 2$ gives the usual spectrum \cite{Strominger:2001pn} \eqref{LowDimSpectrum}, \eqref{LowDimSpectrumSpinJ} for primary states.   For descendants of order $n$ the energy is that of a primary plus $n / r_{{\rm boundary}S^2}$, and for ordinary massive particle states (i.e.~for particle states above the unitarity bound) the descendants are all linearly independent. Now, the conformal characters for massive particles are
\def\chp{3}
\bel{
 \chi\lll{\rm massive}
(\bha)
\equiv \sum_{\vec{n}_\mu}  \expp{- \bha\, \sum_\mu~n_\mu} \ ,
}
where the sum is over energy-raising combinations of conformal generators
$\prod_{\mu = 1}\uu 3 P_\mu^{n_\mu}$ and the dimensionless quantity $\bha$ is
\bel{
\bha \equiv  \frac\beta
{r\lll{{\rm boundary~}S\uu 2}} \ . 
}
Hence, for boundary geometry $S^1 \times S^2$, the partition function for massive single particle states is
\bel{
Z\lll{{{\rm massive~single-}\atop{\rm particles}}} = \sum\lll J \cc 
Z\lll{{{\rm massive~single-}\atop{{\rm particle,~spin~}J}}},
}
with
\algnl{
  Z\lll{{{\rm massive~single-}\atop{{\rm particle,~spin~}0}}} 
&= 2 \cc \chi\lll{\rm massive} 
%\bigg 
(\bha 
%\bigg
\label{FQJEq0}
 ) \cdot \exp{ \left  [ -{{3\bha}\over 2} \right ] }\sum\lll {m\sqd} \cos \bha\sqrt{ m\sqd\ell\lll{dS}\sqd - {9\over 4}},\\
Z\lll{{{\rm massive~single-}\atop{{\rm particle,~spin~}J\geq 1}}} 
&=  
% STARTT THIRD CHANGES MARKED IN BLUE % HEER
{ 2\cc (2J+1) \cc \chi\lll{\rm massive} 
%\bigg 
(\bha 
%\bigg
 ) \cdot \exp{ \left  [ -\bha \Delta(m\sqd \ell\lll{dS}\sqd, J) \right ] }  }
 % ENDD THIRD CHANGES MARKED IN BLUE % HEER
 \label{FQJGr1}
}

At high temperatures, the partition function for any finite collection of bulk particle species is
dominated by the sum over their bulk momenta, as encoded in the CFT
by the conformal characters $\chi(\bha)$.  The
sum over numbers $n\_\mu$ of energy-raising conformal generators
$P_\mu$ can be approximated by an integral, and the behavior of the character is $\chi_{\rm massive}(\bha)   \simeq \bha^{-\chp}$.
For any bulk particle species, or any finite collection thereof, the single-particle
partition function increases monotonically at high temperatures, as
\bel{
Z\lll{{{\rm massive}\atop{{\rm single-particle}}}}  \simeq \sum_{m,J} (2J+1)~\bha\uu{-\chp} \ .
\label{HITempA}
}

When the hierarchy between the dS scale and the scale of new physics is large, $\ell\_{dS} \gg 1/\Lambda$, then in the low temperature limit $\bha\gg 1$ the partition function is well-approximated by a sum over the low-lying spectrum \eqref{LowDimSpectrum} only.\footnote{In the AdS case, this is the limit where the partition function is dominated by a thermal gas of light particles in AdS rather than the AdS/Schwarzschild solution. 
 }  
  The 
expression \eqref{HITempA} increases with increasing temperature.  
For $O(1)$ values of $\bha$ we anticipate the possibility of a transition
to a saddle point that is not globally a quotient of de Sitter space, analogous to the Hawking-Page transition in AdS/CFT.  We shall see candidate geometries for the bulk saddle points controlling such a transition in section \ref{EinsteinSec}.
If this transition can be interpreted as domination of the ensemble
by particle states of large
dimension (of order $\ell \Lambda$) with large statistical degeneracy, then the unitarity
properties of the partition function proposed in this section can be expected to continue to hold, even above the transition.  

If so, then the free energy for each type of particle then increases
monotonically as in \eqref{HITempA}, and the free energy for the CFT as a whole  increases monotonically as well, modulo
issues of uniformity of convergence of the sum over particle species.  We note that
such a behavior would agree with the high-temperature behavior of the contour prescription we shall describe in Section 5.1 rather than that of Section 5.2.

\section{Einstein dS$_4$ and future topology} \label{EinsteinSec}

In this section we consider the WdW wave function for Einstein gravity as a function of the topology at future infinity.  We will compute the actions of complex instantons which solve Einstein's equations with a positive cosmological constant and find a qualitatively similar behaviour to that described above for Vasiliev theory.  

We will compute the WdW wave function in the semi-classical limit.  
The dominant contributions  come from solutions to the equation of motion, which give stationary phase contributions to the wave function.  One can then compute the (appropriately regularized) action of these solutions as a function of boundary data at future infinity.  In principle one should also define some initial conditions in this path integral, which determine the choice of state.  In practice, these initial conditions are defined by a choice of contour for the path integral. 
In general the geometries which appear may be complex.  For example, the standard Hartle-Hawking contour involves geometries which are real in Euclidean signature, but might be complex in Lorentzian signature.  
%We will therefore consider complex solutions to the equations with the appropriate boundary conditions at future infinity.  

An additional complication appears because, in principle, only those solutions which lie on the appropriate stationary phase contour will appear as saddle point contributions to the wave function.   As we do not know how to precisely define the path integral of quantum gravity, the simplest strategy is to assume that all solutions to the equations of motion contribute.   This will be the approach of Section 5.1, where a family of solutions is presented.  The result will be an explosion at high genus similar to that found in Vasiliev theory.  In Section 5.2 we will consider an alternate prescription, where the wave function is computed by an appropriate analytic continuation from Euclidean AdS.  This amounts to a choice of contour which excludes half of the solutions considered in Section 5.1.  This contour does not match the result from Vasiliev theory, and appears to have certain features which are in tension with the criteria discussed in the previous 
section.  We will argue that this choice of contour is unlikely to correctly compute WdW wave function.

\subsection{Complex solutions of dS$_4$ gravity}
\label{sec:complex}

We wish to find solutions of Einstein gravity with a positive cosmological constant which asymptote to $S^1$ 
times a Riemann surface. Our argument essentially follows that of \cite{Anninos:2012ft}, who considered  the case of $S^2 \times S^1$.  See also similar recent computations of \cite{Castro:2012gc, Hartle:2012tv}.

We consider the following metrics
\bel{
\d s^2 = -\frac{\d\tau^2}{f(\tau)}+f(\tau)\d\lambda^2 +\tau^2 \d\Omega^2_\kappa,
}
with % the time-dependent function
\bel{
f(\tau) = \tau^2-\kappa +\frac{\alpha}{\tau}.
}
Here $\d\Omega^{2}_{\kappa}$ is the metric on $S^2$, $T^2$, $\Sigma_g$ for $\kappa=1,0,-1$ respectively, and 
$\lambda$ is an $S^1$ coordinate and is taken to be periodic with period $\lambda_0$. It is easy to check that these solve the equations of motion.  We have set the de Sitter radius $\ell_{dS}$ to 1.  As we are interested in complex solutions, we will not assume at this point that the parameters in this solution are real.  

%
%In general, then, one can seek solutions which are complex The question of which solutions contribute is then really the question of which solutions lie on the appropriate contour of integration in field space.   Real solutions in Lorentzian solution, of course, are expected to describe semi-classical (coherent) states.  In general, one expects that complex solutions to the equations of motion describe semi- from complexified geometries, are expected to describe other 

The geometries are analytic continuations of AdS black holes, where the horizon is an Einstein manifold of positive, zero or negative curvature (see for instance \cite{Emparan:1999pm}). These solutions possess an asymptotic region which approaches the future timelike infinity of dS$_4$ with topology  $S^2 \times S^1$, $T^2 \times S^1$ or $\Sigma_g \times S^1$ respectively. 
%  The question is Also, since we have no way of determining if those complex saddles should lie or not in the contour integration for Einstein gravity, we just treat all these solutions as contributing to the wave function. 

%

If we analytically continue the radius $\ell_{dS}\rightarrow i \ell_{AdS}$ and the bulk coordinate $\tau\rightarrow i r $ then the metric is simply that of an asymptotically AdS black hole in Euclidean signature. The horizon of the AdS black hole is the location in the bulk where the $S^1$ cycle shrinks to zero size smoothly.  For the dS solutions, we will take the location in the bulk where the $S^1$ shrinks to zero size to be at $\tau=\tau_0$.  Regularity of the geometry at this point fixes $\alpha=\kappa \tau_0-\tau_0^3$ and sets the period $\lambda_0$ %of the coordinate $\lambda$ 
as a function of $\tau_0$, % the same relationship between $\lambda_0$ and $\tau_0$,
\bel{
  \lambda_0=\pm \frac{4\pi i \tau_0}{3 \tau_0^2 -\kappa}.
}
%For the dS geometry, $\tau_0$ is a lower bound on the bulk coordinate.

 %Our solutions are such geometries. 
 %They essentially generalize the results of \cite{Anninos:2012ft} to different topologies. 
We now compute the Einstein gravity action, including an appropriate %, comprised of the Einstein-Hilbert part, together with a 
boundary term %chosen to ensure nice variational properties,
\bel{
  S_{L} = \frac{1}{16 \pi G_N}\left[ \int_{\mathcal{M}} \d^{4}x \sqrt{-g} \left( R - \frac{6}{L^2} \right) + 2 \int_{\partial \mathcal{M}}\d^{3}x \sqrt{\gamma} K \right].
}
We are interested in the renormalized action, so we introduce a cutoff surface at $\tau_c$ on which we enforce the boundary metric
\bel{
\gamma_{ij}\d x^{i}\d x^{j}= \tau_{c}^2 \left( \beta^2 \d\theta^2 +\d\Omega^{2}_{\kappa} \right).
}
Here, $\theta$ is a coordinate of $S^1$ with period $2\pi$. The parameter $\beta$ is the radius of the circle at future infinity, which can be regarded as an inverse temperature of the Euclidean boundary theory. For our solutions, we must match the periodicities of $\lambda$ and $\theta$ at $\tau_c$, so
\bel{
\lambda_0=\frac{\tau_c \beta}{\sqrt{f(\tau_c)}}=\frac{\beta}{\sqrt{1-\tfrac{\kappa}{\tau_c^2}}}\left( 1 - \frac{\alpha}{2\tau_c^3} +\cdots \right),
}
where we have neglected terms that fall off as we take our cutoff surface to future infinity. For our solutions in this limit we have
\bel{
\tau_0=\pm \frac{2 \pi i}{3 \beta} \left( -1 \pm \sqrt{1- \tfrac{3 \kappa \beta^2}{4 \pi^2}} \right) +\cdots.
\label{tau0}}

We now compute the action of our solutions. We just need to evaluate the trace of the extrinsic curvature at the boundary
\bel{
K= -\sqrt{f}\left( \frac{f'}{2 f}+ \frac{2}{\tau} \right)\Bigg\vert_{\tau=\tau_c}.
}
The action of our solution is
\bel{
i S_{L}=i \frac{\mathrm{Vol}(\Omega_{\kappa}) \lambda_0}{8 \pi G_N}\left[ -2 \tau_c (\tau_c^2-1) + \frac{\tau_0}{2} (\tau_0^2 -3 \kappa) \right].
}
We can rewrite this in terms of $\beta$ as 
\bel{
i S_{L}=i \frac{\mathrm{Vol}(\Omega_{\kappa})}{8 \pi G_N}\frac{\beta}{\sqrt{1-\tfrac{\kappa}{\tau_c^2}}}\left[  -2 \tau_c (\tau_c^2-1) - \frac{\tau_0}{2} (\tau_0^2 + \kappa) +\cdots \right].
\label{sis}
}
The cutoff-dependent terms diverge as we take $\tau_c\to\infty$; these terms will be removed by a boundary counterterm.

%Before taking temperature limits of our renormalized action we comment on the complex nature of our instantons. 
Each of the solutions described above will lead to a contribution of the form $\Psi = e^{i S_L}$ to the WdW wave function.
Note that the cutoff-dependent terms are oscillatory and do not contribute to $|\Psi|^2$.   We are interested in the renormalized action, which includes only the $\tau_0$-dependent terms in (\ref{sis}). For real $\beta$, $\tau_0$ is complex, indicating that our solutions are complex. Thus $S_L$ is complex and  $|\Psi|^2\neq 1$.

To understand the behaviour of the wave function as a function of genus, let us consider various limits of the periodicity $\beta$.  For $\beta \gg 2\pi$,
\bel{
\tau_0=\pm \sqrt{\frac{\kappa}{3}} \pm \frac{2 \pi i}{3 \beta},
}
and 
\bel{
i S_L= \pm i \left(\frac{\kappa^{3/2} \beta }{12 \sqrt{3} \pi G_N}-\frac{\sqrt{\kappa}\pi}{6\sqrt{3}G_N \beta}\right)\mathrm{Vol}(\Omega_{\kappa}) \pm \left(\frac{\kappa }{12 G_N}-\frac{2 \pi^2}{27 G_N \beta^2}\right)\mathrm{Vol}(\Omega_{\kappa}) + O(\beta^{-3}).
}
For $\beta \ll 2 \pi$ we have two types of solutions
\bel{
\label{action1}
\tau_0= \pm \left( \frac{4 \pi i}{3 \beta} -\frac{i \kappa \beta}{4 \pi} \right),\quad \tau_0=\pm \frac{i \kappa \beta}{4 \pi}
}
and
\bel{
\label{action2}
i S_L= \pm \left( \frac{4 \pi^2 }{27 G_N \beta^2} -\frac{\kappa }{6 G_N} + \frac{ \kappa^2 \beta^2}{64 \pi^2 G_N}\right)\mathrm{Vol}(\Omega_{\kappa}) + O(\beta^3),\quad  i S_L= \pm \frac{\kappa^2 \mathrm{Vol}(\Omega_{\kappa}) \beta^2}{64 G_N \pi^2} + O(\beta^3).
}
It is important to note that the different choices of signs in (\ref{action1}) and (\ref{action2}) correspond to different complex solutions to the equations of motion.  

For the remainder of this subsection we will  assume that all solutions to the equations of motion contribute, so that both possible $\pm$ signs are included.  
With this prescription, our action leads to a 
 wave function $|\Psi|^2$ which diverges in the high temperature ($\beta \to 0$) limit.  This was observed for  $S^2\times S^1$  in \cite{Anninos:2012ft}; we see here that it is a general feature for all topologies.   
 At low temperature ($\beta \to \infty$) we see that the norm of $\Psi$ diverges as well, unlike in the $S^1 \times S^2$ case.  It is tempting to speculate that this is an Einstein gravity version of the infinity of ground states in $Sp(N)$ Vasiliev theory observed in equation (\ref{eq:result}). 

We now address the question of the relative weighting for different topologies.  When comparing different topologies, we must decide at which point in the moduli space of geometries at fixed topologies to evaluate the wave function.  Our choice should be local and conformally invariant; the most natural such choice is to take the curvature of the two dimensional surface to be fixed, so that $\mathrm{Vol}(\Omega_k)$ is proportional to $2(g-1)$.  With this choice, the wave function  increases exponentially with genus.  This is exactly as in the Vasiliev gravity case; the only difference is that, as our computation is based on semi-classical bulk gravity, there is a factor of $1/G_N$ in the exponent rather than $1/G_N^2$.
However, at high temperatures, the Vasiliev result described in section 2.3 scales like $1/G_N$; thus at high temperature the Einstein gravity result matches that of Vasiliev gravity up to an order one constant.
%Thus larger genera are favored without bound. 
%This is an indication that either our computation was too na\"ive, or that probabilities given by the WdW wave function are peaked at larger genera. In any case, this is a puzzle which should lie at the heart of constructing the dS/CFT correspondence, and we will mention some speculations about it in the discussion section.

\subsection{A Euclidean AdS contour}
\label{sec:real}

In the previous section, all solutions to the equations of motion were included in the wave function.  An alternate prescription --- one which is motivated by the similarities between AdS/CFT and dS/CFT --- is to include only those contributions which arise upon analytic continuation from Euclidean AdS space.  This was first discussed in \cite{Maldacena:2002vr}.  

The central observation is that any Euclidean solution of Einstein gravity with a negative cosmological constant will, upon setting $\ell_{AdS} =i\ell_{dS}$, give a (possibly complex) solution to the equations of motion with a positive cosmological constant.  This is because the Euclidean AdS action $S_E$ and the Lorentzian dS action $S_L$ are related by the analytic continuation $S_E(\ell_{AdS}) = i S_L(\ell_{dS})$.  If the original solution has a Euclidean AdS boundary at $z\to0$, then the analytically continued solution will have a Lorentzian dS boundary at $\eta= i z\to 0$.    In Euclidean AdS there is a natural choice of saddles to include in the path integral: those which describe real, smooth metrics in asymptotically AdS space.  The remarkable observation of \cite{Maldacena:2002vr} is that, at the level of perturbation theory around global AdS (dS) space, these solutions analytically continue to those which define the Hartle-Hawking (i.e. Bunch-Davies) state in de Sitter space.

Although the observation of \cite{Maldacena:2002vr} was only made at the level of perturbation theory, it is natural to conjecture that it defines an appropriate contour even at the non-perturbative level. Then the wave function can be computed as follows.  Each smooth, real asymptotically AdS geometry gives a contribution to the Euclidean AdS gravity partition function of the form $Z_{AdS}=e^{-S_E(\ell_{AdS})}$.  The WdW wave function in dS gravity is then computed by taking $\ell_{AdS}=i\ell_{dS}$, so that $\Psi_{dS} = e^{-S_E(i \ell_{dS})}$ where $S_E$ is the action of Euclidean AdS gravity.  The result is summed over all smooth, real solutions of AdS gravity.  It is important to note that, although each such solutions can be interpreted as complex solutions of dS gravity, not all complex solutions of Lorentzian dS gravity will appear in the sum; some of the Lorentzian solutions give geometries which are singular in Euclidean AdS.  This prescription selects a set of preferred  saddles in Lorentzian dS.

In the present case, it is easy to understand which of the saddles constructed in Section 5.1 will contribute to the wave function.  They are  those which correspond to smooth, Euclidean AdS black holes.  The time coordinate of the dS solutions continues to the AdS radial coordinate via $\tau=ir$.  The smooth Euclidean black hole is given by the geometry where the radial coordinate pinches off at the value\footnote{We focus here on the case $\kappa\ne 1$.  When $\kappa=1$ there is in addition a "small black hole" saddle where it caps off at $r_- = {2 \pi - \sqrt{4 \pi^2 - 3 \kappa \beta^2} \over 3 \beta}$.}
\bel{
r_+ = {2 \pi + \sqrt{4 \pi^2 - 3 \kappa \beta^2} \over 3 \beta}.
}
This should be compared to the Lorentzian solutions of equation (\ref{tau0}), where various additional signs are allowed.  We see that only one particular Lorentzian solution continues to the Euclidean AdS black hole; the others continue to singular geometries where the Euclidean horizon does not pinch off smoothly.

The regularized Euclidean action can now be computed, giving (writing $\ell_{AdS} = i \ell_{dS}$)
\bel{
-S_E = \frac{\beta \text{Vol}(\Omega_\kappa) \ell^2_{dS}}{16\pi G_N}\left(-r_+^3+\kappa r_+\right).
} 
The wave function is $\Psi = e^{-S_E}$. We will focus on the limit $\beta \ll 2\pi$, where
\bel{
-S_E= - \left( \frac{4 \pi^2 }{27 G_N \beta^2} -\frac{ \kappa }{6 G_N} + \frac{ \kappa^2 \beta^2}{64 \pi^2 G_N}\right)\mathrm{Vol}(\Omega_{\kappa}) \ell^2_{dS} + O(\beta^3).
}

This result has two important qualitative features.  First, the wave function vanishes in the large genus limit for fixed $\beta$.
Second, for fixed $g$ the wave function vanishes as $\beta \to 0$.
Both of these results are the exact opposite of the behaviour found in Vasiliev theory.  

One possible interpretation of this result is that  Vasiliev gravity is profoundly different from Einstein gravity.  A second interpretation is that, despite its  elegance, the Euclidean AdS contour prescription considered in this section is incorrect.  In particular, a wave function which vanishes as $\beta\to 0$ seems difficult to realize in a theory which exhibits only complex-energy nonunitarity.  This is easiest to see in the boundary CFT, where the wave function with $S^1 \times \Sigma$ asymptotics equals the finite temperature partition function ${\rm Tr}~ e^{-\beta H}$, where $H$ is the generator of Euclidean translations along the circle.  Even though the boundary theory is nonunitary, this partition function still diverges at $\beta\to 0$ in the usual way.  Given this, it seems likely that the WdW wave function cannot be computed simply by analytic continuation of classical instanton actions from Euclidean AdS.\footnote{One possibility is that the Euclidean AdS contour prescription is correct, but 
that one needs to include more than just classical instanton actions.  It might be that once quantum effects are included the wave function would no longer vanish at $\beta \to 0$.  This would signal a breakdown of the semi-classical expansion at high temperature.  This is essentially what was found in \cite{Castro:2012gc}, where the wave function diverged at $\beta \to 0$  in three dimensional gravity.  This was due to a one-loop effect which dominated the naive classical result.  We see no indication that a similar effect happens in the present case, but it would be interesting to investigate this further.}

\section{Discussion} \label{DiscussionSec}

We have found evidence for an instability toward more complicated topology in Vasiliev dS/CFT and have not found a satisfying way of eliminating it.   Einstein calculations indicate a similar trend, albeit with a different strength.

The current understanding of the AdS/CFT duality in the Vasiliev context strongly indicates that the higher spin gravitational fields form an open sector with coupling $1/N$ interacting with a topological closed string sector with coupling $1/N^2$ \cite{Aharony:2011jz, Chang:2012kt, AHJVFuture}.   The Chern-Simons sector in the CFT corresponds to this closed string sector.  A full understanding of the behavior of the dS theory on higher topologies will require an understanding of the dS version of the closed string sector.

In analyzing the meaning of the divergence and possible ways to evade it,
we have found it helpful to distinguish various respects
in which the partition function of a conformal field theory could in principle
deviate from the restricted form satisfied
by a unitary CFT. It is notable that the CS-matter theory and the CFT dual of the
Einstein-gravity limit display an extremely
constrained and non-generic type of nonunitarity in their partition functions, despite the fact that the two CFT live in very different regimes of their coupling constants and anomalous dimensions.  On this basis it is tempting to speculate 
that the positive integrality of the spectral density and the
boundedness-below of the real parts of the energies may be exact properties
%that should be 
satisfied by all conformal field theories that realize any kind of quantum-gravitational theory of de Sitter space holographically, whether it is closely approximated by Einstein gravity or not.

We emphasize that the Einstein gravity computation relies on a crucial assumption, which is that the complex metrics described above lie on an appropriate contour of integration through the space of metrics.  Without a better understanding of the path integral of quantum gravity it is difficult to say whether this is the case.  

One possible implication is that quantum gravity in de Sitter space is intrinsically unstable; this is consistent with other observations about de Sitter gravity \cite{Goheer:2002vf, Kachru:2003aw}.  Another possibility is that de Sitter gravity only makes sense only if we impose a future boundary condition on the WdW wave function, including one which restricts the topology of the spatial slice.   

Yet another possibility is that the divergence need not be interpreted at all, because it describes only probabilities for quantities inaccessible to
any given observer.  The spatial topology of the Universe at future infinity cannot be deduced from inside the causal horizon of any single timelike worldline\footnote{Even in asymptotically flat space, the topology of space is in general unobservable if one assumes the null energy condition \cite{Friedman:1993ty}.}, and
therefore defines not an ``observable'' but a ``meta-observable.''  Such meta-observables are the basic quantities calculated by the dS/CFT framework, or for that matter any theory of the WdW wave function.  However it is well known that quantum mechanics generally does not predict sensible probabilities for physical quantities that are inaccessible to any observer due to the presence of a horizon.  It may be that only suitably coarse-grained probabilities, averaged over all quantities outside a single observer's causal region, will produce an internally consistent set of relative probabilities.  This would require one to incorporate a new rule into the dS/CFT framework that implements observer complementarity as a selection principle for finite quantities.  It is not clear at present how to do so.

\section{Acknowledgements}
We are grateful to O.~Aharony,  A.~Castro, S.~Giombi, D.~Harlow, J.~Maltz, D. Marolf  and especially D.~Anninos for useful discussions. ALJ would like to thank the Stanford Institute for Theoretical Physics (SITP)  for support, where part of this work was completed.  The work of \DJ{}R was supported by SITP and an NSF Graduate Research Fellowship.  The research of SB  and SS is supported by NSF grant 0756174 and the SITP.  The work of SH was supported by World Premier International Research Center Initiative (WPI Initiative), MEXT, Japan, and also by a Grant-in-Aid for Scienti�c Research (22740153) from the Japan Society for Promotion of Science (JSPS). AM, ALJ and AB  are supported by the National Science and Engineering Research Council of Canada.

\end{document}